\documentclass[reprint,prl,twocolumn,showpacs,superscriptaddress,aps]{revtex4-1}
\usepackage[utf8]{inputenc}

\usepackage{graphicx}
\DeclareGraphicsExtensions{.png,.jpg,.eps}

\usepackage{float}
\usepackage{xcolor}
\usepackage{amsmath}
\usepackage{caption}
\usepackage{subcaption}
\usepackage[printwatermark]{xwatermark}
\usepackage{float}

\begin{document}

\title{Controlled 2D Ferromagnetism in \emph{1T}-CrTe$_2$. The role of charge density wave and strain}

\author{Adolfo O. Fumega}
  \email{adolfo.otero.fumega@usc.es}
  \author{Jan Phillips}
\author{Victor Pardo}
  \email{victor.pardo@usc.es}
\affiliation{Departamento de F\'{i}sica Aplicada,
  Universidade de Santiago de Compostela, E-15782 Campus Sur s/n,
  Santiago de Compostela, Spain}
\affiliation{Instituto de Investigaci\'{o}ns Tecnol\'{o}xicas,
  Universidade de Santiago de Compostela, E-15782 Campus Sur s/n,
  Santiago de Compostela, Spain}  
  
\begin{abstract}

Transition metal dichalcogenides are promising candidates to show long-range ferromagnetic order in the single-layer limit.
Based on \emph{ab initio} calculations, we report the emergence of a charge density wave (CDW) phase in monolayer \emph{1T}-CrTe$_2$. We demonstrate that this phase is the ground state in the single-layer limit at any strain value. We obtain an optical phonon mode of $1.96$ THz that connects CDW phase with the undistorted \emph{1T} phase.
Localization of the $a_{1g}$ orbital of CrTe$_2$ produces an out-of-plane orientation of the magnetic moments, circumventing the restrictions of the Mermin-Wagner theorem and producing ferromagnetic long-range order in the two-dimensional limit. This orbital-localization is enhanced by the CDW phase. Tensile strain also increases the localization of this orbital driving the system to become ordered.
CrTe$_2$ becomes an example of a material where the CDW phase produces the stabilization of the long-range ferromagnetic order. 
Our results show that both strain and phase switching are mechanisms to control the 2D ferromagnetic order of CrTe$_2$.

\end{abstract}

\maketitle

\section{Introduction}

The discovery of graphene in 2004 \cite{Graphene} prompted new research areas focused on developing purely two-dimensional (2D) materials that could show emergent physical phenomena and lead to new applications\cite{growth2d, 2dfuelcell,Cao2018}.  
In the last decade, ordered phases such as ferroelectricity or ferromagnetism, that typically occur in bulk, were also reported in 2D materials \cite{2Dferroelectric, 2D_CrGeTe}. 
In particular, the study of 2D ferromagnets results nowadays attractive due to the influence that it could have in other active research areas: spintronics, experiments on quantum anomalous Hall effect, tunneling magneto resistance or spin valves are some examples\cite{2dfm_aplications}.

In order to achieve a 2D ferromagnet, it is necessary to overcome the constraints imposed by the Mermin-Wagner theorem \cite{merminwagner}. It postulates the appearance of gapless spin excitations at finite temperature in isotropic short-range-interaction models, like the isotropic Heisenberg model in 1 or 2D, thus, forbidding long-range ferromagnetic order. In two dimensions the system must lower its symmetry (break spin-rotation invariance) to become a ferromagnet. This can be done by applying an external field or force such as a magnetic field or strain. But, more interestingly, it can also be achieved intrinsically by a strong magnetic anisotropy in the crystal, driven by spin-orbit coupling (SOC). In this study we will see that the emergence of a charge density wave (CDW) phase can also produce the same effect. 

There are already a few examples of systems where ferromagnetism has been observed in the few layer or monolayer limit. These are, e.g. Cr$_2$Ge$_2$Te$_6$ \cite{2D_CrGeTe}, Fe$_3$GeTe$_2$ \cite{fegete}, CrI$_3$ \cite{cri3_2015} and FePS$_3$\cite{feps3_2016}. 
Among them, the family of transition metal dichalcogenides (TMD) has been proposed to be an ideal platform to study 2D ferromagnetism \cite{C7NR06473J,VS2_FM}. These have layered structures with layers bonded via van der Waals interactions (hence, they can be exfoliated down to the monolayer limit),  \cite{xu_ultrathin_2013} and the transition metal in the structure provides the \emph{d}-electrons that in some cases leads to a non-zero magnetic moment.  
Apart from that, several TMD develop a CDW phase  \cite{TiSe2_CDW, TMD_CDW,VSe2_CDW1,Bianco2019,VTe2_CDW,Chen2017} at low temperatures. This phase can deeply influence the magnetic\cite{Fumega2019} and superconductive\cite{Manzeli2017} properties of these compounds. This interplay is yet to be fully understood. In this work, we provide an example of a situation where the CDW phase leads to an enhancement of magnetism (at least in the 2D limit), contrary to, e.g. VSe$_2$ where the opposite occurs\cite{Fumega2019,feng2018electronic,coelho2019charge}.

\begin{figure}[!h]
  \centering
  \includegraphics[width=0.5\textwidth]%
    {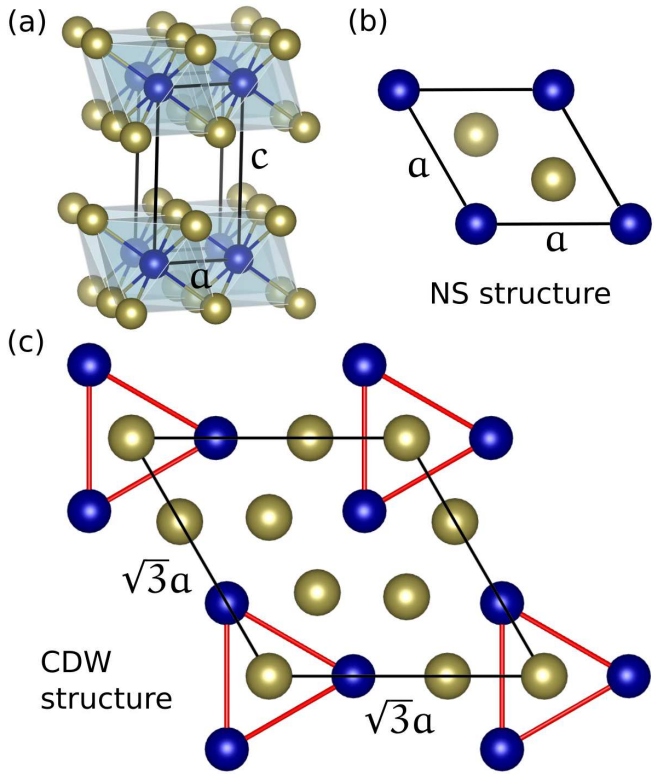}
     \caption{CrTe$_2$ structures used in this study. Cr (Te) atoms in blue (gold). (a) Unit cell in bulk. CrTe$_2$ layers are bonded via van der Waals interactions. Cr atoms octahedrally coordinated by Te atoms. The off-plane direction is a trigonal axis. (b) Top view of monolayer NS (normal state) structure, the single unit cell is depicted by the black rhombus. It contains just 1 Cr and 2 Te atoms. (c) Top view of monolayer CDW structure, a $\sqrt{3}\times\sqrt{3}$ supercell is required to describe its unit cell (black rhombus). Groups of three Cr atoms are formed by a stretching of their bond length (red triangles). }\label{structs_mono}
\end{figure}

This study will be focused on CrTe$_2$, a TMD that crystallizes in a \emph{1T}-phase, in which the Cr-rich hexagonal planes are sandwiched by Te-rich layers. CrTe$_2$ layers of this form are stacked via van der Waals interactions (as shown in Fig. \ref{structs_mono}a). This compound shows a room temperature ferromagnetic phase with a Curie temperature of 310 K and the magnetic moments aligned parallel to the layers \cite{Freitas_2015}. The transition above room temperature makes CrTe$_2$ a highly interesting material from the technological point of view. Based on Density Functional Theory (DFT) calculations, we will analyze below engineering strategies for the bulk structure, such as strain, to align the magnetic moments perpendicular to van der Waals planes. Additionally, we will study in detail the monolayer limit. Our calculations predict the formation of a CDW in the 2D limit. We will analyze the influence that such a phase has to induce long-range ferromagnetic order in 2D. We will also discuss the role of strain at the monolayer level. Our study will show that strain is a possible way to control the ferromagnetic order of CrTe$_2$ in the 2D limit.

\section{Computational methods}
We have performed \emph{ab initio} electronic structure calculations based on DFT\cite{HK,KS} using an all-electron full potential code ({\sc wien2k}\cite{WIEN2k}) on CrTe$_2$.
The exchange-correlation term used for the bulk and monolayer structures was the generalized gradient approximation (GGA) in the Perdew-Burke-Ernzerhof\cite{PBE} scheme. Note that in the case of TMD's van der Waals schemes are required to analyse structural properties\cite{diego2020phonon}. However, our structural analysis focuses on the monolayer limit, where these schemes do not have an effect.
These calculations were performed with a converged k-mesh and a value of  R$_{mt}$K$_{max}$=7.0. The R$_{mt}$ values used were 2.39 and 2.50 in a.u. for Cr and Te respectively. 

The harmonic phonon spectrum of monolayer CrTe$_2$ was computed using the real-space supercell approach\cite{phonopy}.
Scanning tunneling microscope (STM) images were computed using a density of $5\times10^{-5}$ $e^-/$\AA$^3$ associated to the constant current. They were represented using the XCrySDen code\cite{xcrysden}. 

In order to compute the magnetic anisotropy energy (\emph{MAE}), spin-orbit coupling (SOC) was introduced in a second variational manner using the scalar relativistic approximation \cite{SOC_Macdonald}. \emph{MAE} is defined as:
\begin{equation}\label{MAE}
    MAE=E_{In}-E_{Out},
\end{equation}
where $E_{In}$ is the energy per metal atom when the magnetization is set along an in-plane direction (parallel to the Cr layers), while $E_{Out}$ is calculated considering the magnetization points  along the out-of-plane direction (perpendicular to the Cr-rich layers).

\section{The bulk structure}

\begin{figure}[!h]
  \centering
  \includegraphics[width=0.5\textwidth]%
    {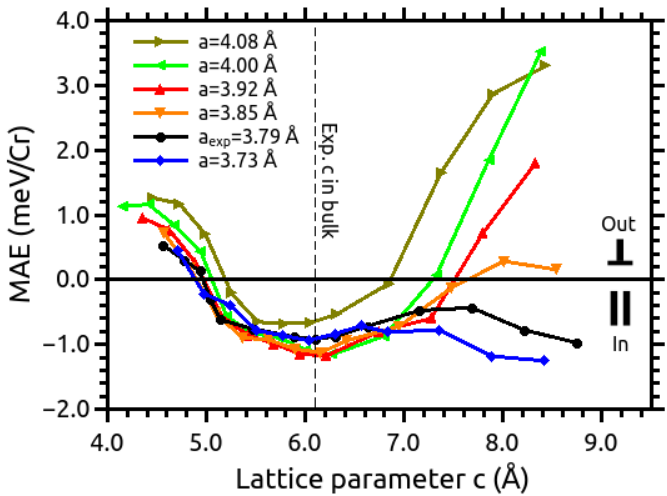}
     \caption{Magnetic anisotropy energy (MAE) as a function of the lattice constant $c$ (out-of-plane) for different lattice constant $a$ (in-plane) values for bulk CrTe$_2$. Positive MAE indicates moments point out of the plane. Decreasing the lattice parameter $c$ drives the magnetic moments to point perpendicular to the CrTe$_2$ layers for any value of $a$ analyzed. Increasing the lattice parameter $c$ drives the magnetic moments to point perpendicular to the Cr layers when the lattice parameter $a$ is also increased compared to the experimental value. }\label{bulk_MAE_c}
\end{figure}

In this section we see how the effect of strain affects the alignment of the magnetic moments. The starting point for the bulk analysis is the experimental structure\cite{Freitas_2015}. We modified the off-plane lattice parameter $c$ and fully relaxed the atomic positions. After this, we introduced SOC for an in-plane magnetization direction and an out-of-plane direction. This allowed us to obtain the needed energies to compute the MAE given by eq. (\ref{MAE}). We then repeated this process for different values of the in-plane lattice parameter $a$, resulting in Fig. \ref{bulk_MAE_c}. Positive values of the MAE in our convention mean that magnetic moments point perpendicular to the Cr layers.  These are of special interest when approaching the 2D limit, where the Mermin-Wagner theorem applies.

The off-plane direction is a trigonal axis for the Te$_6$ octahedron surrounding the Cr cations. Stretching (compressing) the octahedra along this axis leads to a splitting of the otherwise degenerate t$_{2g}$ triplet into a lower(higher-)-lying doublet (so-called $e_{g}^*$) and a higher(lower-)-lying singlet (a$_{1g}$).\cite{khomskii2014book}
Our calculations show that for the experimental values of the lattice parameters, the magnetization points in the plane, which is in agreement with experiment \cite{Freitas_2015}. When
trigonal strain is applied, Fig. \ref{bulk_MAE_c} shows that: i) when the Cr layers come closer to each other (smaller lattice parameter $c$ values), the $MAE$ turns positive. This means that with a high enough uni-axial pressure that approximates Cr layers together, bulk CrTe$_2$ will present an out-of-plane magnetization direction. ii) In a similar way, when approaching the monolayer limit, as Cr layers separate, the magnetic moments also tend to point perpendicular to these, but only for larger lattice parameter $a$ values. In fact, without it, the magnetic moments will stay pointing in-plane (negative values of the $MAE$). 


As discussed before, a magnetization pointing out of the plane is a requirement for ferromagnetic long-range order to exist at a finite temperature in the two-dimensional limit. Thus, it is desirable to understand how to force the magnetization to point out of the plane in the single-layer limit. We will compare below these results in the bulk with similar ones obtained in the 2D limit.

\section{Structural characterization of the monolayer} 

\begin{figure}[!h]
  \centering
  \includegraphics[width=0.5\textwidth]%
    {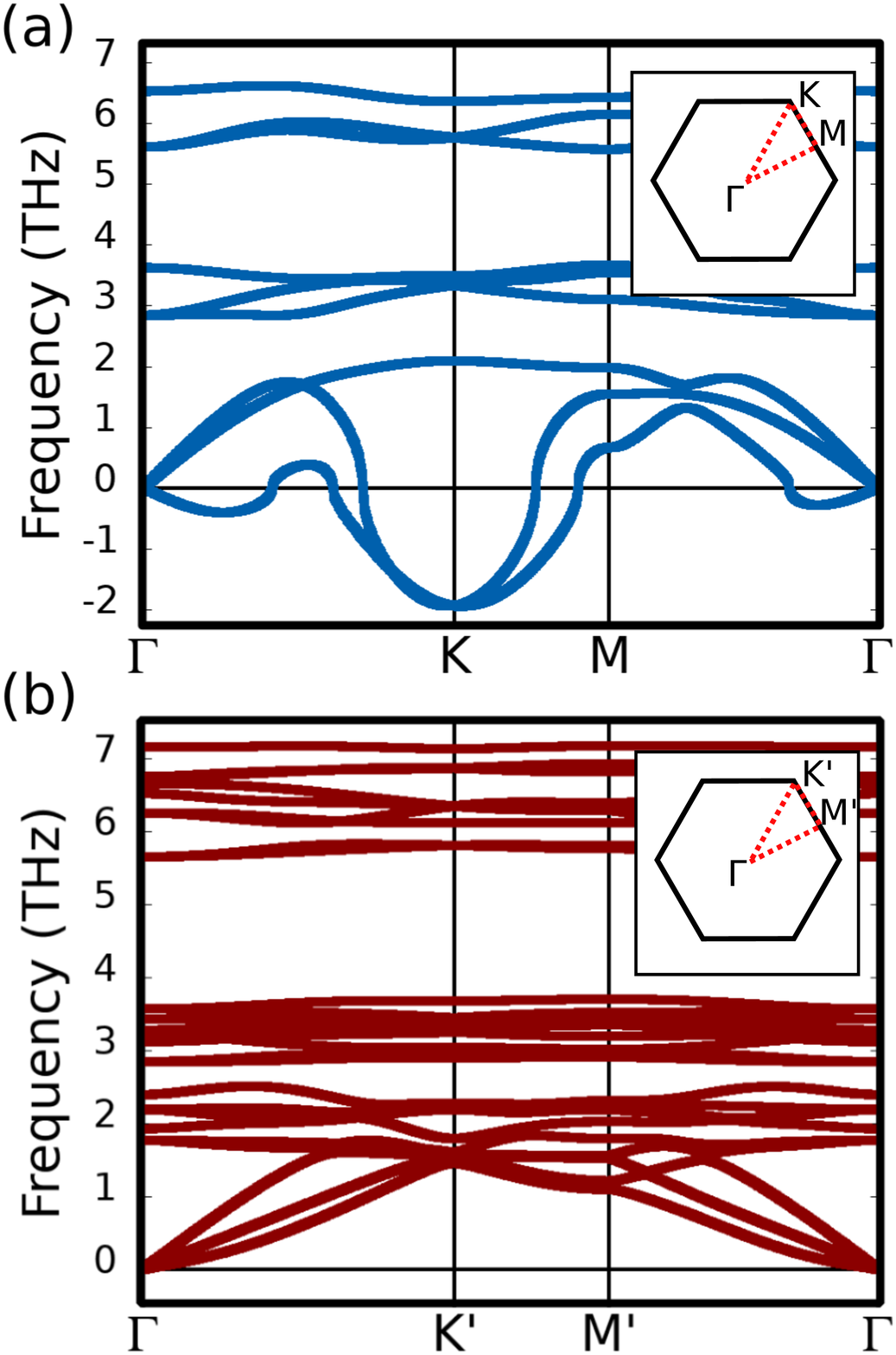}
     \caption{CrTe$_2$ phonon band structures in the monolayer limit. (a) For the NS unit cell structural instabilities appear around the K point in the form of imaginary phonon modes. (b) For the CDW unit cell no instabilities are observed, this is a dynamically stable structure. Brillouin zones along high symmetry paths are included in the insets.}\label{phonon}
\end{figure}

Based on previous studies that report the emergence or enhancement of a CDW phase in some TMD's when the 2D limit is reached \cite{TiSe2_CDW, TMD_CDW,VSe2_CDW1,Bianco2019,VTe2_CDW, Chen2017}, we have analyzed if a CDW state emerges in monolayer CrTe$_2$. To do so, we have performed phonon band structure calculations in what we have defined as the normal state (NS) structure of the monolayer. Figure \ref{structs_mono}b shows the NS structure and its unit cell. This NS structure has a perfect hexagonal arrangement of the Cr atoms as in bulk (space group no. 164), but with the Te atoms fully relaxed in the monolayer limit. In this state, each Cr atom has 6 Cr neighbours at the same distance and the unit cell can be reduced to the one depicted in Fig. \ref{structs_mono}b as a black rhombus. If a CDW phase is the ground state of the system, the phonon band structure must show unstable modes at certain $q$-points related to the supercell structure in which the CDW is described  \cite{Mazin_cdw_phonon}. Figure \ref{phonon}a shows the phonon band structure for NS monolayer CrTe$_2$. A clear instability is observed at the K point ($1/3$,$1/3$) in reciprocal space, which suggests a $\sqrt{3}\times\sqrt{3}$ supercell to describe the CDW state. Note that an instability in the harmonic spectrum is just an indication (not a proof) of the existence of a CDW phase. The introduction of anharmonicities could lead the NS to stabilize, i.e. quenching the appearance of the CDW phase\cite{Bianco2019}. 
Therefore, without including anharmonic effects, the route that we have to follow to demonstrate the existence of the CDW in monolayer is to find a modulated structure lower in energy than the NS structure that presents stable phonon modes.

\begin{figure}[!h]
  \centering
  \includegraphics[width=0.5\textwidth]%
    {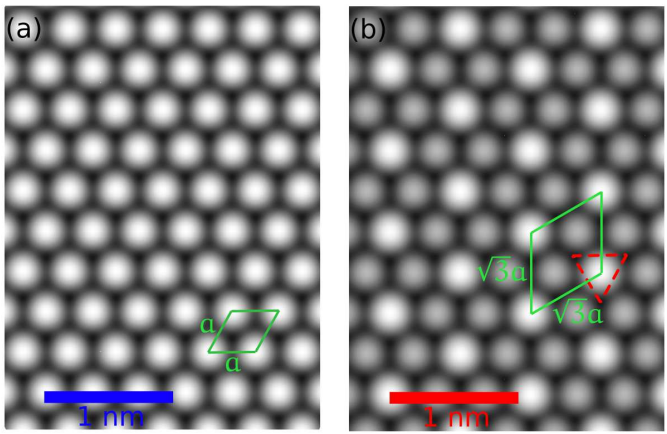}
     \caption{Computed STM images for monolayer CrTe$_2$ (a) in the NS and (b) in the CDW phase. Each white dot is associated to a Te atom in the top layer that forms the Te-Cr-Te monolayer. Unit cells for each phase can be identified (green rhombuses). In the CDW image, an enhancement in the intensity of some Te atoms is produced due to the Cr-bonds stretching (red triangle in Fig. \ref{structs_mono}).}\label{stm}
\end{figure}

Motivated by the harmonic instability at the K point, we have generated a $\sqrt{3}\times\sqrt{3}$ supercell as the one shown in Fig. \ref{structs_mono}c. We have optimized all the atomic positions in it. The result is a reconstruction of the atomic distances. Compared to the NS structure, in which each Cr atom has 6 equidistant Cr neighbours, in the CDW structure each Cr has two short and 4 long Cr-Cr bonds, i.e. a triangular modulation like the one shown in red in Fig. \ref{structs_mono}c appears in the structure. Modulations like this due to the CDW are known to produce new peaks in the X-Ray diffraction pattern\cite{williams_charge_1976}. Nevertheless, in 2D materials this experimental technique does not allow to measure the structural reconstruction that happens. Therefore, we have computed the constant current STM images for the NS (Fig. \ref{stm}a) and the CDW (Fig. \ref{stm}b) structures. 
In these images, the white dots are associated to the top-layer Te atoms that form the Te-Cr-Te sandwich of the 2D material. For the NS case, all of the Te atoms present the same intensity. However, in the CDW case, some of the white dots are brighter and bigger. These dots are the ones associated to the Te atoms that lie inside the red triangle in Fig. \ref{structs_mono}c. The stretching of the Cr atoms forming the triangle make these Te atoms come out of the plane. Thus, producing an increase in the intensity and size observed in the STM image. A symmetry analysis of the CDW structure allows us to identify that the structure corresponds to the space group no. 157, which is a subgroup of the NS structure symmetry (space group no. 164). Note that this subgroup has no inversion symmetry along the z-direction. Thus, this CDW breaks inversion symmetry, just like the Janus monolayers, but without imposing substitutional doping to one of the Te sub-layers\cite{Lu2017, Maghirang2019}.

\begin{figure}[!h]
  \centering
  \includegraphics[width=0.5\textwidth]%
    {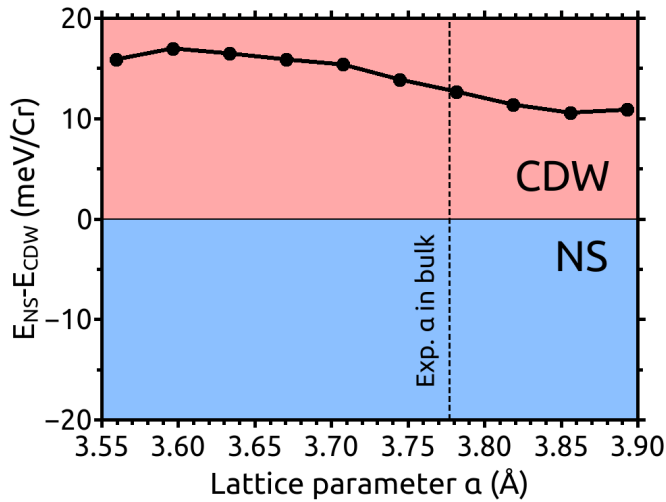}
     \caption{Energy difference between the NS structure and the CDW structure as a function of the $a$ lattice parameter for monolayer CrTe$_2$. Positive values in this difference imply that the CDW phase is more stable than the NS.}\label{energ_mono}
\end{figure}

By now, we have reported the appearance of a CDW in a $\sqrt{3}\times\sqrt{3}$ supercell for monolayer CrTe$_2$. However, we have not demonstrated yet that it corresponds to a ground state at low temperatures. To do so, we have computed the total energies of both the NS and the CDW structures using DFT calculations. Note that now that we have already identified the subgroup associated to the CDW phase, we have reoptimized all the atomic positions and lattice parameters while fixing that space subgroup of the unit cell. This allows to determine which structure is energetically preferred at 0 K. Figure \ref{energ_mono} shows the energy difference between the NS and the CDW structure as a function of the lattice parameter $a$. The experimental value of $a$ in bulk\cite{Freitas_2015} is plotted for comparison.
\emph{Ab initio} calculations show that in the monolayer limit the optimized lattice parameter $a$ is 2\% smaller than the experimental value in bulk, within the uncertainty range for a DFT calculation. Consequently, the overall conclusions obtained for the experimental value are the same as the ones for the optimized value.
We observe that the CDW phase is more stable than the NS at any lattice parameter studied. In particular, for the experimental case, the energy difference is $\sim14$ meV/Cr atom. 
We have also computed the phonon band structure (Fig. \ref{phonon}b) for the CDW structure. We can see that this structure does not show dynamic instabilities in the form of imaginary phonon modes in the phonon band structure, and hence it provides the ground state for monolayer CrTe$_2$. Note that the labels of the reciprocal space in Fig. \ref{phonon}b have been primed to emphasize that the first Brillouin zone is reduced in the CDW phase.
Moreover, we have identified an optical mode in the CDW structure that drives this phase to the NS one. This mode has a frequency of 1.96 THz and could be activated to switch between both phases. 

\section{Ferromagnetism in the 2D limit}

Now that we have already seen that at low temperature a CDW phase emerges on monolayer CrTe$_2$, we can analyze its effect on the ferromagnetic order of the compound. But before doing that, we are going to see the effect that it produces on the electronic structure.

\begin{figure}[!h]
  \centering
  \includegraphics[width=0.5\textwidth]%
    {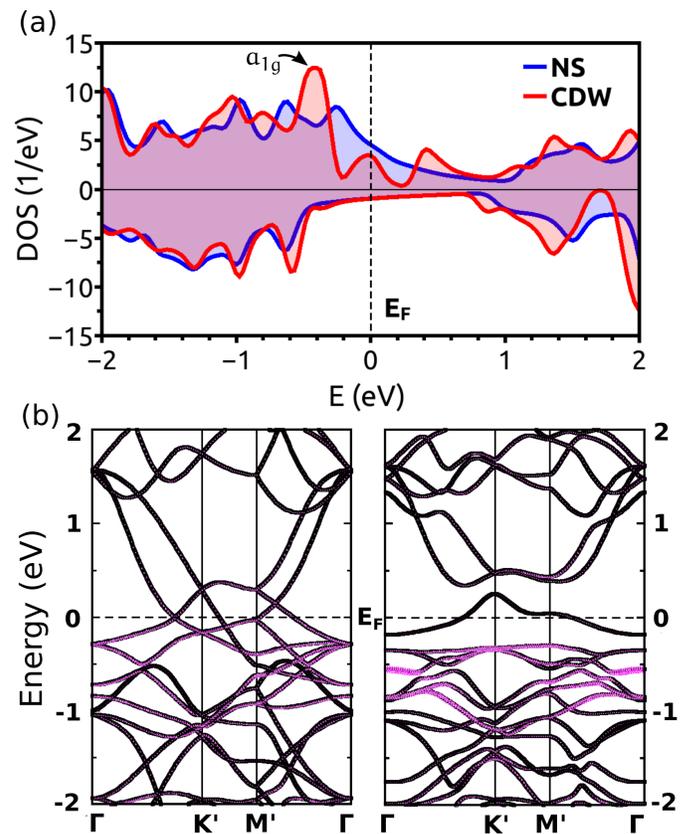}
     \caption{(a) Density of states (DOS) as a function of energy for the NS structure in blue and the CDW one in red. The majority (minority) spin channel is represented as positive (negative). (b) Band structures for the majority spin channels. Left (right) panel for the NS (CDW) structure. The $a_{1g}$ orbital character is highlighted in pink.}\label{dos_bands_mono}
\end{figure}

Typically, the occurrence of a CDW phase leads to the opening of electronic gaps or pseudogaps in the vicinity of the Fermi level \cite{terashima_charge-density_2003,PhysRevLett.100.196402,Chen2017}. 
In order to see the rearrangement that the electronic structure of CrTe$_2$ undergoes within the CDW phase, we have computed the electronic density of states (DOS) and the band structures for both phases. Figure \ref{dos_bands_mono}a shows the DOS for the NS (CDW) phase in blue (red) near the Fermi level.  
As discussed above for the bulk case, the octahedral environment around the Cr atoms is trigonally distorted and hence the $d$-orbitals split into 3 sets of energy levels, one $a_{1g}$ singlet and two ($e_g$ and $e_{g}^*$) doublets, the latter being split from the cubic t$_{2g}$ triplet. The occupation of the $a_{1g}$ and two electrons in the wider $e_{g}^*$ bands in the majority spin channel leads to a total magnetic moment of 3 $\mu_B/$Cr, corresponding to an approximate Cr$^{3+}$ valence, but with a high degree of hybridization with Te-$p$ orbitals, which produces the observed metallic state. The minority spin channel presents roughly no occupied Cr d states (apart from some bonding character in the Te p bands).
In Fig. \ref{dos_bands_mono}a, in the majority channel, we can see that gaps are opened around the Fermi level as a consequence of the CDW (shown in red).
This can be analyzed in more detail in the band structure (Fig. \ref{dos_bands_mono}b) in which the majority spin channels are represented. On the left (right) panel, the NS (CDW) band structure is shown, both computed in the $\sqrt{3}\times\sqrt{3}$ supercell. Due to the atomic rearrangement produced by the CDW, bands hybridize  opening a gap below and above the Fermi level that leaves a flat single band crossing the Fermi level. This depletion of states around the Fermi level reduces the total energy of the system, making the CDW structure the ground state solution. 
We can also see that the $a_{1g}$ orbital, whose character is represented in pink in Fig. \ref{dos_bands_mono}b, gets more localized. This can also be identified as a peak in the DOS just below the Fermi level for the CDW case (Fig. \ref{dos_bands_mono}a).

\begin{figure}[!h]
  \centering
  \includegraphics[width=0.5\textwidth]%
    {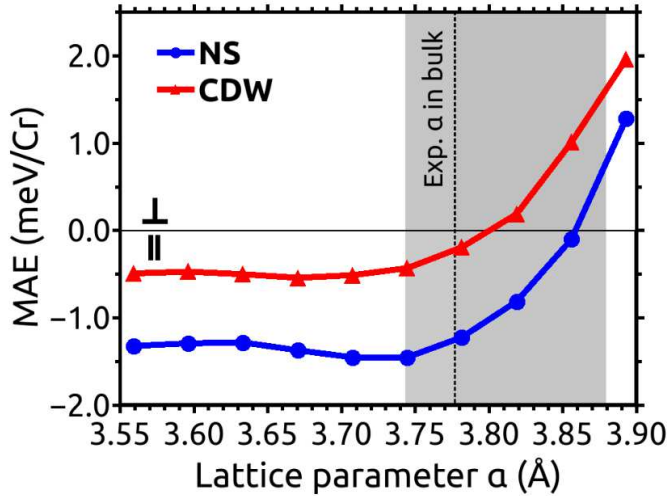}
     \caption{$MAE$ as a function of the lattice parameter $a$ for 2D CrTe$_2$. The CDW phase and tensile strain drive the magnetic moments to point perpendicular to the monolayer. In the shaded area we could control the ferromagnetic order via strain or phase transition.}\label{soc_mono_strain}
\end{figure}

Regarding the magnetic properties of the system, we have already stated that monolayer CrTe$_2$ presents a magnetization of  3 $\mu_B$ per Cr atom. 
In the bulk case, the system is a room-temperature ferromagnet in which the in-plane magnetic exchange couplings (the ones relevant in the 2D limit) are ferromagnetic. Considering previous works in which the dimensionality effect is studied for similar Cr van der Waals compounds \cite{Wang_2011,OFumega2020}, and given the small changes in the Cr-Cr distance between monolayer and bulk limits, the in-plane magnetic exchange couplings will remain ferromagnetic in monolayer CrTe$_2$.
The question that arises now is if the moments are long-range ordered in the 2D limit, thus providing a ferromagnetic state. 
In orther to check this, we have computed the $MAE$ as given by eq. (\ref{MAE}). Fig. \ref{soc_mono_strain} shows the $MAE$ as a function of strain for the NS and the CDW phases. An out-of-plane $MAE$ (positive values in our convention) circumvents the restrictions of the Mermin-Wagner theorem leading to long-range ferromagnetic order to be stable at finite temperatures.

In Fig \ref{soc_mono_strain} we observe that for the experimental value of the in-plane lattice constant, a long-range FM order is on the verge of being feasible. A very small tensile strain produces a spin reorientation for the CDW phase with the solution with an out-of-plane magnetization being stable. A substantially larger tensile strain is required to produce the same effect in the NS structure. It can be seen that the CDW phase has a higher tendency to have the magnetic moments aligned perpendicular to the layer, i.e. every $MAE$ value for the CDW is more positive (or less negative) than the corresponding one in the NS. This tendency for the moments to orient out of plane is accompanied of a stronger moment localization in the CDW phase, as we can see in the band structure analyzed above. When tensile strain is applied, the states also become more localized. 
Bandwidths are reduced by an increase of the inter-atomic bond length which causes a reduction of the in-plane hopping. In this respect, tensile strain operates in the same direction as the CDW does.

We saw above that for CrTe$_2$ in the bulk,  strain is also a mechanism responsible for producing an out-of-plane orientation of the magnetization in this compound. In particular, enlarging the in-plane lattice parameter (reducing the Cr d bandwidth, and producing larger moment localization) increases the tendency for the magnetic moments to point out of the plane.

This kind of out-of-plane to in-plane magnetic moment reorientation has been observed in similar (bulk) van der Waals ferromagnets like Cr$_2$Ge$_2$Te$_6$ when pressure is applied \cite{spin_anys_press}. The effect of pressure goes in the opposite direction than tensile strain. It reduces the lattice parameter, thus increasing the bandwidth (or delocalization) of the $d$ orbitals and hence driving the moments parallel to the layers. Thus, this relationship between magnetocrystalline anisotropy and moment orientation in trigonally distorted Cr$^{3+}$-based 2D magnets is somewhat general.

We observe that for the experimental value of the lattice parameter $a$, ferromagnetic order is on the verge of stability in the CDW phase. Small values of strain could lead the system to become ordered, and vice versa. Thus, strain provides an effective way to control 2D ferromagnetism in monolayer CrTe$_2$. Apart from this mechanism, it can also be controlled by activating the 1.96 THz phonon mode that we have identified in the previous section. This would drive the CDW phase to the NS one, thus quenching the magnetic order. The shaded area in grey in Fig. \ref{soc_mono_strain} provides the range of lattice parameter values in which we could control the ferromagnetic order of 2D CrTe$_2$.

\section{Summary and conclusions}

In this article we have analyzed the magnetic and structural properties of CrTe$_2$.

In the first part we focused on the bulk structure. We saw how different distortions applied to the lattice parameters could drive the moments to point perpendicular to the van der Waals layers. By decreasing the lattice parameter $c$ an increase in the hopping between layers is produced. This leads the moments to point out-of-plane. By increasing the lattice parameter $c$, i.e. increasing the distance between layers and approaching the 2D limit, magnetic moments point perpendicular if the lattice parameter $a$ is also increased. 

A detailed description of the monolayer limit was presented in the following sections. First, a structural analysis was performed. We found that a CDW  associated to the symmetry space group no. 157 emerges at low temperatures. This CDW phase corresponds to the ground state at any strain value, and it helps eliminating the dynamic instabilities that appear in the phonon spectrum for the NS structure. We showed computed STM images of NS and CDW phases to guide experimentalists to identify each phase. We have also determined that an optical phonon mode of $1.96$ THz is responsible for the transition between phases. Thus, prompting a route to switch between them in a controlled way \cite{Nova1075}. 
The analysis of the electronic structure showed that the CDW phase opens pseudogaps around the Fermi level and localizes the $d$ orbitals. This is found to drive the magnetic moments to point perpendicular to the monolayer, thus, overcoming the constrains imposed by the Mermin-Wagner theorem and producing long-range ferrromagnetic order in the 2D limit. Tensile strain is also found to increase the localization of the $d$ orbitals and hence produce ferromagnetism in both phases. This mechanism of moment localization associated to the magnetization pointing out of plane operates in other Cr$^{3+}$-based low-dimensional magnets.

In this article, we have highlighted the importance that the structure has in providing a good description of the magnetic properties in CrTe$_2$. Opposite to what has been reported for other similar systems, such as VSe$_2$, \cite{Fumega2019} where a CDW is incompatible with a long-range FM order, we have found that the emergence of a CDW drives CrTe$_2$ to become a ferromagnet in the single-layer limit. Our work provides additional understanding for the interplay between different competing phases in low-dimensional materials and gives fundamental insight in order to find 2D ferromagnetism in TMD's and related systems.

\section{Supporting information}
A video  showing the phonon mode of 1.96 THz that connects the CDW phase and the NS one has been included as supplemental material.
Monolayer CDW and NS cif files have also been included.

\section*{Acknowledgements}
This work is supported by the MINECO of Spain through the project PGC2018-101334-B-C21. A.O.F. thanks MECD for the financial support received through the FPU grant FPU16/02572. We made use of the facilities provided by the Galician Supercomputing Center (CESGA). We also thank S. Blanco-Canosa for fruitful discussions.


\begin{thebibliography}{44}%
\makeatletter
\providecommand \@ifxundefined [1]{%
 \@ifx{#1\undefined}
}%
\providecommand \@ifnum [1]{%
 \ifnum #1\expandafter \@firstoftwo
 \else \expandafter \@secondoftwo
 \fi
}%
\providecommand \@ifx [1]{%
 \ifx #1\expandafter \@firstoftwo
 \else \expandafter \@secondoftwo
 \fi
}%
\providecommand \natexlab [1]{#1}%
\providecommand \enquote  [1]{``#1''}%
\providecommand \bibnamefont  [1]{#1}%
\providecommand \bibfnamefont [1]{#1}%
\providecommand \citenamefont [1]{#1}%
\providecommand \href@noop [0]{\@secondoftwo}%
\providecommand \href [0]{\begingroup \@sanitize@url \@href}%
\providecommand \@href[1]{\@@startlink{#1}\@@href}%
\providecommand \@@href[1]{\endgroup#1\@@endlink}%
\providecommand \@sanitize@url [0]{\catcode `\\12\catcode `\$12\catcode
  `\&12\catcode `\#12\catcode `\^12\catcode `\_12\catcode `\%12\relax}%
\providecommand \@@startlink[1]{}%
\providecommand \@@endlink[0]{}%
\providecommand \url  [0]{\begingroup\@sanitize@url \@url }%
\providecommand \@url [1]{\endgroup\@href {#1}{\urlprefix }}%
\providecommand \urlprefix  [0]{URL }%
\providecommand \Eprint [0]{\href }%
\providecommand \doibase [0]{http://dx.doi.org/}%
\providecommand \selectlanguage [0]{\@gobble}%
\providecommand \bibinfo  [0]{\@secondoftwo}%
\providecommand \bibfield  [0]{\@secondoftwo}%
\providecommand \translation [1]{[#1]}%
\providecommand \BibitemOpen [0]{}%
\providecommand \bibitemStop [0]{}%
\providecommand \bibitemNoStop [0]{.\EOS\space}%
\providecommand \EOS [0]{\spacefactor3000\relax}%
\providecommand \BibitemShut  [1]{\csname bibitem#1\endcsname}%
\let\auto@bib@innerbib\@empty
\bibitem [{\citenamefont {Novoselov}\ \emph {et~al.}(2004)\citenamefont
  {Novoselov}, \citenamefont {Geim}, \citenamefont {Morozov}, \citenamefont
  {Jiang}, \citenamefont {Zhang}, \citenamefont {Dubonos}, \citenamefont
  {Grigorieva},\ and\ \citenamefont {Firsov}}]{Graphene}%
  \BibitemOpen
  \bibfield  {author} {\bibinfo {author} {\bibfnamefont {K.~S.}\ \bibnamefont
  {Novoselov}}, \bibinfo {author} {\bibfnamefont {A.~K.}\ \bibnamefont {Geim}},
  \bibinfo {author} {\bibfnamefont {S.~V.}\ \bibnamefont {Morozov}}, \bibinfo
  {author} {\bibfnamefont {D.}~\bibnamefont {Jiang}}, \bibinfo {author}
  {\bibfnamefont {Y.}~\bibnamefont {Zhang}}, \bibinfo {author} {\bibfnamefont
  {S.~V.}\ \bibnamefont {Dubonos}}, \bibinfo {author} {\bibfnamefont {I.~V.}\
  \bibnamefont {Grigorieva}}, \ and\ \bibinfo {author} {\bibfnamefont {A.~A.}\
  \bibnamefont {Firsov}},\ }\href {\doibase 10.1126/science.1102896} {\bibfield
   {journal} {\bibinfo  {journal} {Science}\ }\textbf {\bibinfo {volume}
  {306}},\ \bibinfo {pages} {666} (\bibinfo {year} {2004})}\BibitemShut
  {NoStop}%
\bibitem [{\citenamefont {Geng}\ and\ \citenamefont {Yang}()}]{growth2d}%
  \BibitemOpen
  \bibfield  {author} {\bibinfo {author} {\bibfnamefont {D.}~\bibnamefont
  {Geng}}\ and\ \bibinfo {author} {\bibfnamefont {H.~Y.}\ \bibnamefont
  {Yang}},\ }\href {\doibase 10.1002/adma.201800865} {\bibfield  {journal}
  {\bibinfo  {journal} {Advanced Materials}\ }\textbf {\bibinfo {volume}
  {30}},\ \bibinfo {pages} {1800865}}\BibitemShut {NoStop}%
\bibitem [{\citenamefont {Khan}\ \emph {et~al.}(2019)\citenamefont {Khan},
  \citenamefont {Tareen}, \citenamefont {Aslam}, \citenamefont {Zhang},
  \citenamefont {Wang}, \citenamefont {Ouyang}, \citenamefont {Gou},\ and\
  \citenamefont {Zhang}}]{2dfuelcell}%
  \BibitemOpen
  \bibfield  {author} {\bibinfo {author} {\bibfnamefont {K.}~\bibnamefont
  {Khan}}, \bibinfo {author} {\bibfnamefont {A.~K.}\ \bibnamefont {Tareen}},
  \bibinfo {author} {\bibfnamefont {M.}~\bibnamefont {Aslam}}, \bibinfo
  {author} {\bibfnamefont {Y.}~\bibnamefont {Zhang}}, \bibinfo {author}
  {\bibfnamefont {R.}~\bibnamefont {Wang}}, \bibinfo {author} {\bibfnamefont
  {Z.}~\bibnamefont {Ouyang}}, \bibinfo {author} {\bibfnamefont
  {Z.}~\bibnamefont {Gou}}, \ and\ \bibinfo {author} {\bibfnamefont
  {H.}~\bibnamefont {Zhang}},\ }\href {\doibase 10.1039/C9NR05919A} {\bibfield
  {journal} {\bibinfo  {journal} {Nanoscale}\ }\textbf {\bibinfo {volume}
  {11}},\ \bibinfo {pages} {21622} (\bibinfo {year} {2019})}\BibitemShut
  {NoStop}%
\bibitem [{\citenamefont {Cao}\ \emph {et~al.}(2018)\citenamefont {Cao},
  \citenamefont {Fatemi}, \citenamefont {Demir}, \citenamefont {Fang},
  \citenamefont {Tomarken}, \citenamefont {Luo}, \citenamefont
  {Sanchez-Yamagishi}, \citenamefont {Watanabe}, \citenamefont {Taniguchi},
  \citenamefont {Kaxiras},\ and\ \citenamefont {et~al.}}]{Cao2018}%
  \BibitemOpen
  \bibfield  {author} {\bibinfo {author} {\bibfnamefont {Y.}~\bibnamefont
  {Cao}}, \bibinfo {author} {\bibfnamefont {V.}~\bibnamefont {Fatemi}},
  \bibinfo {author} {\bibfnamefont {A.}~\bibnamefont {Demir}}, \bibinfo
  {author} {\bibfnamefont {S.}~\bibnamefont {Fang}}, \bibinfo {author}
  {\bibfnamefont {S.~L.}\ \bibnamefont {Tomarken}}, \bibinfo {author}
  {\bibfnamefont {J.~Y.}\ \bibnamefont {Luo}}, \bibinfo {author} {\bibfnamefont
  {J.~D.}\ \bibnamefont {Sanchez-Yamagishi}}, \bibinfo {author} {\bibfnamefont
  {K.}~\bibnamefont {Watanabe}}, \bibinfo {author} {\bibfnamefont
  {T.}~\bibnamefont {Taniguchi}}, \bibinfo {author} {\bibfnamefont
  {E.}~\bibnamefont {Kaxiras}}, \ and\ \bibinfo {author} {\bibnamefont
  {et~al.}},\ }\href {\doibase 10.1038/nature26154} {\bibfield  {journal}
  {\bibinfo  {journal} {Nature}\ }\textbf {\bibinfo {volume} {556}},\ \bibinfo
  {pages} {80} (\bibinfo {year} {2018})}\BibitemShut {NoStop}%
\bibitem [{\citenamefont {Chang}\ \emph {et~al.}(2016)\citenamefont {Chang},
  \citenamefont {Liu}, \citenamefont {Lin}, \citenamefont {Wang}, \citenamefont
  {Zhao}, \citenamefont {Zhang}, \citenamefont {Jin}, \citenamefont {Zhong},
  \citenamefont {Hu}, \citenamefont {Duan},\ and\ \citenamefont
  {et~al.}}]{2Dferroelectric}%
  \BibitemOpen
  \bibfield  {author} {\bibinfo {author} {\bibfnamefont {K.}~\bibnamefont
  {Chang}}, \bibinfo {author} {\bibfnamefont {J.}~\bibnamefont {Liu}}, \bibinfo
  {author} {\bibfnamefont {H.}~\bibnamefont {Lin}}, \bibinfo {author}
  {\bibfnamefont {N.}~\bibnamefont {Wang}}, \bibinfo {author} {\bibfnamefont
  {K.}~\bibnamefont {Zhao}}, \bibinfo {author} {\bibfnamefont {A.}~\bibnamefont
  {Zhang}}, \bibinfo {author} {\bibfnamefont {F.}~\bibnamefont {Jin}}, \bibinfo
  {author} {\bibfnamefont {Y.}~\bibnamefont {Zhong}}, \bibinfo {author}
  {\bibfnamefont {X.}~\bibnamefont {Hu}}, \bibinfo {author} {\bibfnamefont
  {W.}~\bibnamefont {Duan}}, \ and\ \bibinfo {author} {\bibnamefont {et~al.}},\
  }\href {\doibase 10.1126/science.aad8609} {\bibfield  {journal} {\bibinfo
  {journal} {Science}\ }\textbf {\bibinfo {volume} {353}},\ \bibinfo {pages}
  {274} (\bibinfo {year} {2016})},\ \Eprint
  {http://arxiv.org/abs/https://science.sciencemag.org/content/353/6296/274.full.pdf}
  {https://science.sciencemag.org/content/353/6296/274.full.pdf} \BibitemShut
  {NoStop}%
\bibitem [{\citenamefont {Gong}\ \emph {et~al.}(2017)\citenamefont {Gong},
  \citenamefont {Li}, \citenamefont {Li}, \citenamefont {Ji}, \citenamefont
  {Stern}, \citenamefont {Xia}, \citenamefont {Cao}, \citenamefont {Bao},
  \citenamefont {Wang}, \citenamefont {Wang},\ and\ \citenamefont
  {et~al.}}]{2D_CrGeTe}%
  \BibitemOpen
  \bibfield  {author} {\bibinfo {author} {\bibfnamefont {C.}~\bibnamefont
  {Gong}}, \bibinfo {author} {\bibfnamefont {L.}~\bibnamefont {Li}}, \bibinfo
  {author} {\bibfnamefont {Z.}~\bibnamefont {Li}}, \bibinfo {author}
  {\bibfnamefont {H.}~\bibnamefont {Ji}}, \bibinfo {author} {\bibfnamefont
  {A.}~\bibnamefont {Stern}}, \bibinfo {author} {\bibfnamefont
  {Y.}~\bibnamefont {Xia}}, \bibinfo {author} {\bibfnamefont {T.}~\bibnamefont
  {Cao}}, \bibinfo {author} {\bibfnamefont {W.}~\bibnamefont {Bao}}, \bibinfo
  {author} {\bibfnamefont {C.}~\bibnamefont {Wang}}, \bibinfo {author}
  {\bibfnamefont {Y.}~\bibnamefont {Wang}}, \ and\ \bibinfo {author}
  {\bibnamefont {et~al.}},\ }\href {\doibase 10.1038/nature22060} {\bibfield
  {journal} {\bibinfo  {journal} {Nature}\ }\textbf {\bibinfo {volume} {546}},\
  \bibinfo {pages} {265} (\bibinfo {year} {2017})}\BibitemShut {NoStop}%
\bibitem [{\citenamefont {Huang}\ \emph {et~al.}(2020)\citenamefont {Huang},
  \citenamefont {Zhang}, \citenamefont {Xu}, \citenamefont {Wang},
  \citenamefont {Zhang},\ and\ \citenamefont {Zhang}}]{2dfm_aplications}%
  \BibitemOpen
  \bibfield  {author} {\bibinfo {author} {\bibfnamefont {P.}~\bibnamefont
  {Huang}}, \bibinfo {author} {\bibfnamefont {P.}~\bibnamefont {Zhang}},
  \bibinfo {author} {\bibfnamefont {S.}~\bibnamefont {Xu}}, \bibinfo {author}
  {\bibfnamefont {H.}~\bibnamefont {Wang}}, \bibinfo {author} {\bibfnamefont
  {X.}~\bibnamefont {Zhang}}, \ and\ \bibinfo {author} {\bibfnamefont
  {H.}~\bibnamefont {Zhang}},\ }\href {\doibase 10.1039/C9NR08890C} {\bibfield
  {journal} {\bibinfo  {journal} {Nanoscale}\ }\textbf {\bibinfo {volume}
  {12}},\ \bibinfo {pages} {2309} (\bibinfo {year} {2020})}\BibitemShut
  {NoStop}%
\bibitem [{\citenamefont {Mermin}\ and\ \citenamefont
  {Wagner}(1966)}]{merminwagner}%
  \BibitemOpen
  \bibfield  {author} {\bibinfo {author} {\bibfnamefont {N.~D.}\ \bibnamefont
  {Mermin}}\ and\ \bibinfo {author} {\bibfnamefont {H.}~\bibnamefont
  {Wagner}},\ }\href {\doibase 10.1103/PhysRevLett.17.1133} {\bibfield
  {journal} {\bibinfo  {journal} {Phys. Rev. Lett.}\ }\textbf {\bibinfo
  {volume} {17}},\ \bibinfo {pages} {1133} (\bibinfo {year}
  {1966})}\BibitemShut {NoStop}%
\bibitem [{\citenamefont {Fei}\ \emph {et~al.}(2018)\citenamefont {Fei},
  \citenamefont {Huang}, \citenamefont {Malinowski}, \citenamefont {Wang},
  \citenamefont {Song}, \citenamefont {Sanchez}, \citenamefont {Yao},
  \citenamefont {Xiao}, \citenamefont {Zhu}, \citenamefont {May},\ and\
  \citenamefont {et~al.}}]{fegete}%
  \BibitemOpen
  \bibfield  {author} {\bibinfo {author} {\bibfnamefont {Z.}~\bibnamefont
  {Fei}}, \bibinfo {author} {\bibfnamefont {B.}~\bibnamefont {Huang}}, \bibinfo
  {author} {\bibfnamefont {P.}~\bibnamefont {Malinowski}}, \bibinfo {author}
  {\bibfnamefont {W.}~\bibnamefont {Wang}}, \bibinfo {author} {\bibfnamefont
  {T.}~\bibnamefont {Song}}, \bibinfo {author} {\bibfnamefont {J.}~\bibnamefont
  {Sanchez}}, \bibinfo {author} {\bibfnamefont {W.}~\bibnamefont {Yao}},
  \bibinfo {author} {\bibfnamefont {D.}~\bibnamefont {Xiao}}, \bibinfo {author}
  {\bibfnamefont {X.}~\bibnamefont {Zhu}}, \bibinfo {author} {\bibfnamefont
  {A.~F.}\ \bibnamefont {May}}, \ and\ \bibinfo {author} {\bibnamefont
  {et~al.}},\ }\href {\doibase 10.1038/s41563-018-0149-7} {\bibfield  {journal}
  {\bibinfo  {journal} {Nature Materials}\ }\textbf {\bibinfo {volume} {17}},\
  \bibinfo {pages} {778} (\bibinfo {year} {2018})}\BibitemShut {NoStop}%
\bibitem [{\citenamefont {Zhang}\ \emph {et~al.}(2015)\citenamefont {Zhang},
  \citenamefont {Qu}, \citenamefont {Zhu},\ and\ \citenamefont
  {Lam}}]{cri3_2015}%
  \BibitemOpen
  \bibfield  {author} {\bibinfo {author} {\bibfnamefont {W.-B.}\ \bibnamefont
  {Zhang}}, \bibinfo {author} {\bibfnamefont {Q.}~\bibnamefont {Qu}}, \bibinfo
  {author} {\bibfnamefont {P.}~\bibnamefont {Zhu}}, \ and\ \bibinfo {author}
  {\bibfnamefont {C.-H.}\ \bibnamefont {Lam}},\ }\href {\doibase
  10.1039/C5TC02840J} {\bibfield  {journal} {\bibinfo  {journal} {J. Mater.
  Chem. C}\ }\textbf {\bibinfo {volume} {3}},\ \bibinfo {pages} {12457}
  (\bibinfo {year} {2015})}\BibitemShut {NoStop}%
\bibitem [{\citenamefont {Lee}\ \emph {et~al.}(2016)\citenamefont {Lee},
  \citenamefont {Lee}, \citenamefont {Ryoo}, \citenamefont {Kang},
  \citenamefont {Kim}, \citenamefont {Kim}, \citenamefont {Park}, \citenamefont
  {Park},\ and\ \citenamefont {Cheong}}]{feps3_2016}%
  \BibitemOpen
  \bibfield  {author} {\bibinfo {author} {\bibfnamefont {J.-U.}\ \bibnamefont
  {Lee}}, \bibinfo {author} {\bibfnamefont {S.}~\bibnamefont {Lee}}, \bibinfo
  {author} {\bibfnamefont {J.~H.}\ \bibnamefont {Ryoo}}, \bibinfo {author}
  {\bibfnamefont {S.}~\bibnamefont {Kang}}, \bibinfo {author} {\bibfnamefont
  {T.~Y.}\ \bibnamefont {Kim}}, \bibinfo {author} {\bibfnamefont
  {P.}~\bibnamefont {Kim}}, \bibinfo {author} {\bibfnamefont {C.-H.}\
  \bibnamefont {Park}}, \bibinfo {author} {\bibfnamefont {J.-G.}\ \bibnamefont
  {Park}}, \ and\ \bibinfo {author} {\bibfnamefont {H.}~\bibnamefont
  {Cheong}},\ }\href {\doibase 10.1021/acs.nanolett.6b03052} {\bibfield
  {journal} {\bibinfo  {journal} {Nano Letters}\ }\textbf {\bibinfo {volume}
  {16}},\ \bibinfo {pages} {7433} (\bibinfo {year} {2016})},\ \Eprint
  {http://arxiv.org/abs/https://doi.org/10.1021/acs.nanolett.6b03052}
  {https://doi.org/10.1021/acs.nanolett.6b03052} \BibitemShut {NoStop}%
\bibitem [{\citenamefont {Du}\ \emph {et~al.}(2017)\citenamefont {Du},
  \citenamefont {Xia}, \citenamefont {Xiong}, \citenamefont {Wang},
  \citenamefont {Jia},\ and\ \citenamefont {Li}}]{C7NR06473J}%
  \BibitemOpen
  \bibfield  {author} {\bibinfo {author} {\bibfnamefont {J.}~\bibnamefont
  {Du}}, \bibinfo {author} {\bibfnamefont {C.}~\bibnamefont {Xia}}, \bibinfo
  {author} {\bibfnamefont {W.}~\bibnamefont {Xiong}}, \bibinfo {author}
  {\bibfnamefont {T.}~\bibnamefont {Wang}}, \bibinfo {author} {\bibfnamefont
  {Y.}~\bibnamefont {Jia}}, \ and\ \bibinfo {author} {\bibfnamefont
  {J.}~\bibnamefont {Li}},\ }\href {\doibase 10.1039/C7NR06473J} {\bibfield
  {journal} {\bibinfo  {journal} {Nanoscale}\ }\textbf {\bibinfo {volume}
  {9}},\ \bibinfo {pages} {17585} (\bibinfo {year} {2017})}\BibitemShut
  {NoStop}%
\bibitem [{\citenamefont {Guo}\ \emph {et~al.}()\citenamefont {Guo},
  \citenamefont {Deng}, \citenamefont {Sun}, \citenamefont {Li}, \citenamefont
  {Zhao}, \citenamefont {Wu}, \citenamefont {Chu}, \citenamefont {Zhang},
  \citenamefont {Pan}, \citenamefont {Zheng},\ and\ \citenamefont
  {et~al.}}]{VS2_FM}%
  \BibitemOpen
  \bibfield  {author} {\bibinfo {author} {\bibfnamefont {Y.}~\bibnamefont
  {Guo}}, \bibinfo {author} {\bibfnamefont {H.}~\bibnamefont {Deng}}, \bibinfo
  {author} {\bibfnamefont {X.}~\bibnamefont {Sun}}, \bibinfo {author}
  {\bibfnamefont {X.}~\bibnamefont {Li}}, \bibinfo {author} {\bibfnamefont
  {J.}~\bibnamefont {Zhao}}, \bibinfo {author} {\bibfnamefont {J.}~\bibnamefont
  {Wu}}, \bibinfo {author} {\bibfnamefont {W.}~\bibnamefont {Chu}}, \bibinfo
  {author} {\bibfnamefont {S.}~\bibnamefont {Zhang}}, \bibinfo {author}
  {\bibfnamefont {H.}~\bibnamefont {Pan}}, \bibinfo {author} {\bibfnamefont
  {X.}~\bibnamefont {Zheng}}, \ and\ \bibinfo {author} {\bibnamefont
  {et~al.}},\ }\href {\doibase 10.1002/adma.201700715} {\bibfield  {journal}
  {\bibinfo  {journal} {Advanced Materials}\ }\textbf {\bibinfo {volume}
  {29}},\ \bibinfo {pages} {1700715}}\BibitemShut {NoStop}%
\bibitem [{\citenamefont {Xu}\ \emph {et~al.}(2013)\citenamefont {Xu},
  \citenamefont {Chen}, \citenamefont {Li}, \citenamefont {Wu}, \citenamefont
  {Guo}, \citenamefont {Zhao}, \citenamefont {Wu},\ and\ \citenamefont
  {Xie}}]{xu_ultrathin_2013}%
  \BibitemOpen
  \bibfield  {author} {\bibinfo {author} {\bibfnamefont {K.}~\bibnamefont
  {Xu}}, \bibinfo {author} {\bibfnamefont {P.}~\bibnamefont {Chen}}, \bibinfo
  {author} {\bibfnamefont {X.}~\bibnamefont {Li}}, \bibinfo {author}
  {\bibfnamefont {C.}~\bibnamefont {Wu}}, \bibinfo {author} {\bibfnamefont
  {Y.}~\bibnamefont {Guo}}, \bibinfo {author} {\bibfnamefont {J.}~\bibnamefont
  {Zhao}}, \bibinfo {author} {\bibfnamefont {X.}~\bibnamefont {Wu}}, \ and\
  \bibinfo {author} {\bibfnamefont {Y.}~\bibnamefont {Xie}},\ }\href {\doibase
  10.1002/anie.201304337} {\bibfield  {journal} {\bibinfo  {journal}
  {Angewandte Chemie International Edition}\ }\textbf {\bibinfo {volume}
  {52}},\ \bibinfo {pages} {10477} (\bibinfo {year} {2013})}\BibitemShut
  {NoStop}%
\bibitem [{\citenamefont {Bussmann-Holder}\ and\ \citenamefont
  {Büttner}(2002)}]{TiSe2_CDW}%
  \BibitemOpen
  \bibfield  {author} {\bibinfo {author} {\bibfnamefont {A.}~\bibnamefont
  {Bussmann-Holder}}\ and\ \bibinfo {author} {\bibfnamefont {H.}~\bibnamefont
  {Büttner}},\ }\href {http://stacks.iop.org/0953-8984/14/i=34/a=316}
  {\bibfield  {journal} {\bibinfo  {journal} {Journal of Physics: Condensed
  Matter}\ }\textbf {\bibinfo {volume} {14}},\ \bibinfo {pages} {7973}
  (\bibinfo {year} {2002})}\BibitemShut {NoStop}%
\bibitem [{\citenamefont {Wilson}\ \emph {et~al.}(1975)\citenamefont {Wilson},
  \citenamefont {Salvo},\ and\ \citenamefont {Mahajan}}]{TMD_CDW}%
  \BibitemOpen
  \bibfield  {author} {\bibinfo {author} {\bibfnamefont {J.}~\bibnamefont
  {Wilson}}, \bibinfo {author} {\bibfnamefont {F.~D.}\ \bibnamefont {Salvo}}, \
  and\ \bibinfo {author} {\bibfnamefont {S.}~\bibnamefont {Mahajan}},\ }\href
  {\doibase 10.1080/00018737500101391} {\bibfield  {journal} {\bibinfo
  {journal} {Advances in Physics}\ }\textbf {\bibinfo {volume} {24}},\ \bibinfo
  {pages} {117} (\bibinfo {year} {1975})},\ \Eprint
  {http://arxiv.org/abs/https://doi.org/10.1080/00018737500101391}
  {https://doi.org/10.1080/00018737500101391} \BibitemShut {NoStop}%
\bibitem [{\citenamefont {Eaglesham}\ \emph {et~al.}(1986)\citenamefont
  {Eaglesham}, \citenamefont {Withers},\ and\ \citenamefont
  {Bird}}]{VSe2_CDW1}%
  \BibitemOpen
  \bibfield  {author} {\bibinfo {author} {\bibfnamefont {D.~J.}\ \bibnamefont
  {Eaglesham}}, \bibinfo {author} {\bibfnamefont {R.~L.}\ \bibnamefont
  {Withers}}, \ and\ \bibinfo {author} {\bibfnamefont {D.~M.}\ \bibnamefont
  {Bird}},\ }\href {http://stacks.iop.org/0022-3719/19/i=3/a=006} {\bibfield
  {journal} {\bibinfo  {journal} {Journal of Physics C: Solid State Physics}\
  }\textbf {\bibinfo {volume} {19}},\ \bibinfo {pages} {359} (\bibinfo {year}
  {1986})}\BibitemShut {NoStop}%
\bibitem [{\citenamefont {Bianco}\ \emph {et~al.}(2019)\citenamefont {Bianco},
  \citenamefont {Errea}, \citenamefont {Monacelli}, \citenamefont {Calandra},\
  and\ \citenamefont {Mauri}}]{Bianco2019}%
  \BibitemOpen
  \bibfield  {author} {\bibinfo {author} {\bibfnamefont {R.}~\bibnamefont
  {Bianco}}, \bibinfo {author} {\bibfnamefont {I.}~\bibnamefont {Errea}},
  \bibinfo {author} {\bibfnamefont {L.}~\bibnamefont {Monacelli}}, \bibinfo
  {author} {\bibfnamefont {M.}~\bibnamefont {Calandra}}, \ and\ \bibinfo
  {author} {\bibfnamefont {F.}~\bibnamefont {Mauri}},\ }\href {\doibase
  10.1021/acs.nanolett.9b00504} {\bibfield  {journal} {\bibinfo  {journal}
  {Nano Letters}\ }\textbf {\bibinfo {volume} {19}},\ \bibinfo {pages} {3098}
  (\bibinfo {year} {2019})}\BibitemShut {NoStop}%
\bibitem [{\citenamefont {Wang}\ \emph {et~al.}(2019)\citenamefont {Wang},
  \citenamefont {Ren}, \citenamefont {Li}, \citenamefont {Wang}, \citenamefont
  {Peng}, \citenamefont {Yu}, \citenamefont {Duan},\ and\ \citenamefont
  {Zhou}}]{VTe2_CDW}%
  \BibitemOpen
  \bibfield  {author} {\bibinfo {author} {\bibfnamefont {Y.}~\bibnamefont
  {Wang}}, \bibinfo {author} {\bibfnamefont {J.}~\bibnamefont {Ren}}, \bibinfo
  {author} {\bibfnamefont {J.}~\bibnamefont {Li}}, \bibinfo {author}
  {\bibfnamefont {Y.}~\bibnamefont {Wang}}, \bibinfo {author} {\bibfnamefont
  {H.}~\bibnamefont {Peng}}, \bibinfo {author} {\bibfnamefont {P.}~\bibnamefont
  {Yu}}, \bibinfo {author} {\bibfnamefont {W.}~\bibnamefont {Duan}}, \ and\
  \bibinfo {author} {\bibfnamefont {S.}~\bibnamefont {Zhou}},\ }\href {\doibase
  10.1103/PhysRevB.100.241404} {\bibfield  {journal} {\bibinfo  {journal}
  {Phys. Rev. B}\ }\textbf {\bibinfo {volume} {100}},\ \bibinfo {pages}
  {241404} (\bibinfo {year} {2019})}\BibitemShut {NoStop}%
\bibitem [{\citenamefont {Chen}\ \emph {et~al.}(2017)\citenamefont {Chen},
  \citenamefont {Pai}, \citenamefont {Chan}, \citenamefont {Takayama},
  \citenamefont {Xu}, \citenamefont {Karn}, \citenamefont {Hasegawa},
  \citenamefont {Chou}, \citenamefont {Mo}, \citenamefont {Fedorov},\ and\
  \citenamefont {et~al.}}]{Chen2017}%
  \BibitemOpen
  \bibfield  {author} {\bibinfo {author} {\bibfnamefont {P.}~\bibnamefont
  {Chen}}, \bibinfo {author} {\bibfnamefont {W.~W.}\ \bibnamefont {Pai}},
  \bibinfo {author} {\bibfnamefont {Y.-H.}\ \bibnamefont {Chan}}, \bibinfo
  {author} {\bibfnamefont {A.}~\bibnamefont {Takayama}}, \bibinfo {author}
  {\bibfnamefont {C.-Z.}\ \bibnamefont {Xu}}, \bibinfo {author} {\bibfnamefont
  {A.}~\bibnamefont {Karn}}, \bibinfo {author} {\bibfnamefont {S.}~\bibnamefont
  {Hasegawa}}, \bibinfo {author} {\bibfnamefont {M.~Y.}\ \bibnamefont {Chou}},
  \bibinfo {author} {\bibfnamefont {S.-K.}\ \bibnamefont {Mo}}, \bibinfo
  {author} {\bibfnamefont {A.-V.}\ \bibnamefont {Fedorov}}, \ and\ \bibinfo
  {author} {\bibnamefont {et~al.}},\ }\href {\doibase
  10.1038/s41467-017-00641-1} {\bibfield  {journal} {\bibinfo  {journal}
  {Nature Communications}\ }\textbf {\bibinfo {volume} {8}},\ \bibinfo {pages}
  {516} (\bibinfo {year} {2017})}\BibitemShut {NoStop}%
\bibitem [{\citenamefont {Fumega}\ \emph {et~al.}(2019)\citenamefont {Fumega},
  \citenamefont {Gobbi}, \citenamefont {Dreher}, \citenamefont {Wan},
  \citenamefont {Gonz{\'a}lez-Orellana}, \citenamefont {Pe{\~n}a-D{\'i}az},
  \citenamefont {Rogero}, \citenamefont {Herrero-Mart{\'i}n}, \citenamefont
  {Gargiani}, \citenamefont {Ilyn},\ and\ \citenamefont {et~al.}}]{Fumega2019}%
  \BibitemOpen
  \bibfield  {author} {\bibinfo {author} {\bibfnamefont {A.~O.}\ \bibnamefont
  {Fumega}}, \bibinfo {author} {\bibfnamefont {M.}~\bibnamefont {Gobbi}},
  \bibinfo {author} {\bibfnamefont {P.}~\bibnamefont {Dreher}}, \bibinfo
  {author} {\bibfnamefont {W.}~\bibnamefont {Wan}}, \bibinfo {author}
  {\bibfnamefont {C.}~\bibnamefont {Gonz{\'a}lez-Orellana}}, \bibinfo {author}
  {\bibfnamefont {M.}~\bibnamefont {Pe{\~n}a-D{\'i}az}}, \bibinfo {author}
  {\bibfnamefont {C.}~\bibnamefont {Rogero}}, \bibinfo {author} {\bibfnamefont
  {J.}~\bibnamefont {Herrero-Mart{\'i}n}}, \bibinfo {author} {\bibfnamefont
  {P.}~\bibnamefont {Gargiani}}, \bibinfo {author} {\bibfnamefont
  {M.}~\bibnamefont {Ilyn}}, \ and\ \bibinfo {author} {\bibnamefont {et~al.}},\
  }\href {\doibase 10.1021/acs.jpcc.9b08868} {\bibfield  {journal} {\bibinfo
  {journal} {The Journal of Physical Chemistry C}\ }\textbf {\bibinfo {volume}
  {123}},\ \bibinfo {pages} {27802} (\bibinfo {year} {2019})}\BibitemShut
  {NoStop}%
\bibitem [{\citenamefont {Manzeli}\ \emph {et~al.}(2017)\citenamefont
  {Manzeli}, \citenamefont {Ovchinnikov}, \citenamefont {Pasquier},
  \citenamefont {Yazyev},\ and\ \citenamefont {Kis}}]{Manzeli2017}%
  \BibitemOpen
  \bibfield  {author} {\bibinfo {author} {\bibfnamefont {S.}~\bibnamefont
  {Manzeli}}, \bibinfo {author} {\bibfnamefont {D.}~\bibnamefont
  {Ovchinnikov}}, \bibinfo {author} {\bibfnamefont {D.}~\bibnamefont
  {Pasquier}}, \bibinfo {author} {\bibfnamefont {O.~V.}\ \bibnamefont
  {Yazyev}}, \ and\ \bibinfo {author} {\bibfnamefont {A.}~\bibnamefont {Kis}},\
  }\href {\doibase 10.1038/natrevmats.2017.33} {\bibfield  {journal} {\bibinfo
  {journal} {Nature Reviews Materials}\ }\textbf {\bibinfo {volume} {2}},\
  \bibinfo {pages} {17033} (\bibinfo {year} {2017})}\BibitemShut {NoStop}%
\bibitem [{\citenamefont {Feng}\ \emph {et~al.}(2018)\citenamefont {Feng},
  \citenamefont {Biswas}, \citenamefont {Rajan}, \citenamefont {Watson},
  \citenamefont {Mazzola}, \citenamefont {Clark}, \citenamefont {Underwood},
  \citenamefont {Markovic}, \citenamefont {McLaren}, \citenamefont {Hunter},\
  and\ \citenamefont {et~al.}}]{feng2018electronic}%
  \BibitemOpen
  \bibfield  {author} {\bibinfo {author} {\bibfnamefont {J.}~\bibnamefont
  {Feng}}, \bibinfo {author} {\bibfnamefont {D.}~\bibnamefont {Biswas}},
  \bibinfo {author} {\bibfnamefont {A.}~\bibnamefont {Rajan}}, \bibinfo
  {author} {\bibfnamefont {M.~D.}\ \bibnamefont {Watson}}, \bibinfo {author}
  {\bibfnamefont {F.}~\bibnamefont {Mazzola}}, \bibinfo {author} {\bibfnamefont
  {O.~J.}\ \bibnamefont {Clark}}, \bibinfo {author} {\bibfnamefont
  {K.}~\bibnamefont {Underwood}}, \bibinfo {author} {\bibfnamefont
  {I.}~\bibnamefont {Markovic}}, \bibinfo {author} {\bibfnamefont
  {M.}~\bibnamefont {McLaren}}, \bibinfo {author} {\bibfnamefont
  {A.}~\bibnamefont {Hunter}}, \ and\ \bibinfo {author} {\bibnamefont
  {et~al.}},\ }\href@noop {} {\bibfield  {journal} {\bibinfo  {journal} {Nano
  letters}\ }\textbf {\bibinfo {volume} {18}},\ \bibinfo {pages} {4493}
  (\bibinfo {year} {2018})}\BibitemShut {NoStop}%
\bibitem [{\citenamefont {Coelho}\ \emph {et~al.}(2019)\citenamefont {Coelho},
  \citenamefont {Nguyen~Cong}, \citenamefont {Bonilla}, \citenamefont
  {Kolekar}, \citenamefont {Phan}, \citenamefont {Avila}, \citenamefont
  {Asensio}, \citenamefont {Oleynik},\ and\ \citenamefont
  {Batzill}}]{coelho2019charge}%
  \BibitemOpen
  \bibfield  {author} {\bibinfo {author} {\bibfnamefont {P.~M.}\ \bibnamefont
  {Coelho}}, \bibinfo {author} {\bibfnamefont {K.}~\bibnamefont {Nguyen~Cong}},
  \bibinfo {author} {\bibfnamefont {M.}~\bibnamefont {Bonilla}}, \bibinfo
  {author} {\bibfnamefont {S.}~\bibnamefont {Kolekar}}, \bibinfo {author}
  {\bibfnamefont {M.-H.}\ \bibnamefont {Phan}}, \bibinfo {author}
  {\bibfnamefont {J.}~\bibnamefont {Avila}}, \bibinfo {author} {\bibfnamefont
  {M.~C.}\ \bibnamefont {Asensio}}, \bibinfo {author} {\bibfnamefont {I.~I.}\
  \bibnamefont {Oleynik}}, \ and\ \bibinfo {author} {\bibfnamefont
  {M.}~\bibnamefont {Batzill}},\ }\href@noop {} {\bibfield  {journal} {\bibinfo
   {journal} {The Journal of Physical Chemistry C}\ }\textbf {\bibinfo {volume}
  {123}},\ \bibinfo {pages} {14089} (\bibinfo {year} {2019})}\BibitemShut
  {NoStop}%
\bibitem [{\citenamefont {Freitas}\ \emph {et~al.}(2015)\citenamefont
  {Freitas}, \citenamefont {Weht}, \citenamefont {Sulpice}, \citenamefont
  {Remenyi}, \citenamefont {Strobel}, \citenamefont {Gay}, \citenamefont
  {Marcus},\ and\ \citenamefont {N{\'{u}}{\~{n}}ez-Regueiro}}]{Freitas_2015}%
  \BibitemOpen
  \bibfield  {author} {\bibinfo {author} {\bibfnamefont {D.~C.}\ \bibnamefont
  {Freitas}}, \bibinfo {author} {\bibfnamefont {R.}~\bibnamefont {Weht}},
  \bibinfo {author} {\bibfnamefont {A.}~\bibnamefont {Sulpice}}, \bibinfo
  {author} {\bibfnamefont {G.}~\bibnamefont {Remenyi}}, \bibinfo {author}
  {\bibfnamefont {P.}~\bibnamefont {Strobel}}, \bibinfo {author} {\bibfnamefont
  {F.}~\bibnamefont {Gay}}, \bibinfo {author} {\bibfnamefont {J.}~\bibnamefont
  {Marcus}}, \ and\ \bibinfo {author} {\bibfnamefont {M.}~\bibnamefont
  {N{\'{u}}{\~{n}}ez-Regueiro}},\ }\href {\doibase
  10.1088/0953-8984/27/17/176002} {\bibfield  {journal} {\bibinfo  {journal}
  {Journal of Physics: Condensed Matter}\ }\textbf {\bibinfo {volume} {27}},\
  \bibinfo {pages} {176002} (\bibinfo {year} {2015})}\BibitemShut {NoStop}%
\bibitem [{\citenamefont {Hohenberg}\ and\ \citenamefont {Kohn}(1964)}]{HK}%
  \BibitemOpen
  \bibfield  {author} {\bibinfo {author} {\bibfnamefont {P.}~\bibnamefont
  {Hohenberg}}\ and\ \bibinfo {author} {\bibfnamefont {W.}~\bibnamefont
  {Kohn}},\ }\href@noop {} {\bibfield  {journal} {\bibinfo  {journal} {Phys.
  Rev.}\ }\textbf {\bibinfo {volume} {136}},\ \bibinfo {pages} {B864} (\bibinfo
  {year} {1964})}\BibitemShut {NoStop}%
\bibitem [{\citenamefont {Kohn}\ and\ \citenamefont {Sham}(1965)}]{KS}%
  \BibitemOpen
  \bibfield  {author} {\bibinfo {author} {\bibfnamefont {W.}~\bibnamefont
  {Kohn}}\ and\ \bibinfo {author} {\bibfnamefont {L.~J.}\ \bibnamefont
  {Sham}},\ }\href@noop {} {\bibfield  {journal} {\bibinfo  {journal} {Phys.
  Rev.}\ }\textbf {\bibinfo {volume} {140}},\ \bibinfo {pages} {A1133}
  (\bibinfo {year} {1965})}\BibitemShut {NoStop}%
\bibitem [{\citenamefont {Schwarz}\ and\ \citenamefont {Blaha}(2003)}]{WIEN2k}%
  \BibitemOpen
  \bibfield  {author} {\bibinfo {author} {\bibfnamefont {K.}~\bibnamefont
  {Schwarz}}\ and\ \bibinfo {author} {\bibfnamefont {P.}~\bibnamefont
  {Blaha}},\ }\href@noop {} {\bibfield  {journal} {\bibinfo  {journal} {Comp.
  Mater. Sci.}\ }\textbf {\bibinfo {volume} {28}},\ \bibinfo {pages} {259}
  (\bibinfo {year} {2003})}\BibitemShut {NoStop}%
\bibitem [{\citenamefont {Perdew}\ \emph {et~al.}(1996)\citenamefont {Perdew},
  \citenamefont {Burke},\ and\ \citenamefont {Ernzerhof}}]{PBE}%
  \BibitemOpen
  \bibfield  {author} {\bibinfo {author} {\bibfnamefont {J.~P.}\ \bibnamefont
  {Perdew}}, \bibinfo {author} {\bibfnamefont {K.}~\bibnamefont {Burke}}, \
  and\ \bibinfo {author} {\bibfnamefont {M.}~\bibnamefont {Ernzerhof}},\
  }\href@noop {} {\bibfield  {journal} {\bibinfo  {journal} {Phys. Rev. Lett.}\
  }\textbf {\bibinfo {volume} {77}},\ \bibinfo {pages} {3865} (\bibinfo {year}
  {1996})}\BibitemShut {NoStop}%
\bibitem [{\citenamefont {Diego}\ \emph {et~al.}(2020)\citenamefont {Diego},
  \citenamefont {Said}, \citenamefont {Mahatha}, \citenamefont {Bianco},
  \citenamefont {Monacelli}, \citenamefont {Calandra}, \citenamefont {Mauri},
  \citenamefont {Rossnagel}, \citenamefont {Errea},\ and\ \citenamefont
  {Blanco-Canosa}}]{diego2020phonon}%
  \BibitemOpen
  \bibfield  {author} {\bibinfo {author} {\bibfnamefont {J.}~\bibnamefont
  {Diego}}, \bibinfo {author} {\bibfnamefont {A.~H.}\ \bibnamefont {Said}},
  \bibinfo {author} {\bibfnamefont {S.~K.}\ \bibnamefont {Mahatha}}, \bibinfo
  {author} {\bibfnamefont {R.}~\bibnamefont {Bianco}}, \bibinfo {author}
  {\bibfnamefont {L.}~\bibnamefont {Monacelli}}, \bibinfo {author}
  {\bibfnamefont {M.}~\bibnamefont {Calandra}}, \bibinfo {author}
  {\bibfnamefont {F.}~\bibnamefont {Mauri}}, \bibinfo {author} {\bibfnamefont
  {K.}~\bibnamefont {Rossnagel}}, \bibinfo {author} {\bibfnamefont
  {I.}~\bibnamefont {Errea}}, \ and\ \bibinfo {author} {\bibfnamefont
  {S.}~\bibnamefont {Blanco-Canosa}},\ }\href@noop {} {\enquote {\bibinfo
  {title} {Phonon collapse and van der waals melting of the 3d charge density
  wave of vse$_2$, arxiv:2007.08413},}\ } (\bibinfo {year} {2020}),\ \Eprint
  {http://arxiv.org/abs/2007.08413} {arXiv:2007.08413 [cond-mat.mtrl-sci]}
  \BibitemShut {NoStop}%
\bibitem [{\citenamefont {Togo}\ and\ \citenamefont {Tanaka}(2015)}]{phonopy}%
  \BibitemOpen
  \bibfield  {author} {\bibinfo {author} {\bibfnamefont {A.}~\bibnamefont
  {Togo}}\ and\ \bibinfo {author} {\bibfnamefont {I.}~\bibnamefont {Tanaka}},\
  }\href@noop {} {\bibfield  {journal} {\bibinfo  {journal} {Scr. Mater.}\
  }\textbf {\bibinfo {volume} {108}},\ \bibinfo {pages} {1} (\bibinfo {year}
  {2015})}\BibitemShut {NoStop}%
\bibitem [{\citenamefont {Kokalj}(1999)}]{xcrysden}%
  \BibitemOpen
  \bibfield  {author} {\bibinfo {author} {\bibfnamefont {A.}~\bibnamefont
  {Kokalj}},\ }\href {\doibase https://doi.org/10.1016/S1093-3263(99)00028-5}
  {\bibfield  {journal} {\bibinfo  {journal} {Journal of Molecular Graphics and
  Modelling}\ }\textbf {\bibinfo {volume} {17}},\ \bibinfo {pages} {176 }
  (\bibinfo {year} {1999})}\BibitemShut {NoStop}%
\bibitem [{\citenamefont {MacDonald}\ \emph {et~al.}(1980)\citenamefont
  {MacDonald}, \citenamefont {Picket},\ and\ \citenamefont
  {Koelling}}]{SOC_Macdonald}%
  \BibitemOpen
  \bibfield  {author} {\bibinfo {author} {\bibfnamefont {A.~H.}\ \bibnamefont
  {MacDonald}}, \bibinfo {author} {\bibfnamefont {W.~E.}\ \bibnamefont
  {Picket}}, \ and\ \bibinfo {author} {\bibfnamefont {D.~D.}\ \bibnamefont
  {Koelling}},\ }\href@noop {} {\bibfield  {journal} {\bibinfo  {journal}
  {Journal of Physics C: Solid State Physics}\ }\textbf {\bibinfo {volume}
  {13}},\ \bibinfo {pages} {2675} (\bibinfo {year} {1980})}\BibitemShut
  {NoStop}%
\bibitem [{\citenamefont {Khomskii}(2014)}]{khomskii2014book}%
  \BibitemOpen
  \bibfield  {author} {\bibinfo {author} {\bibfnamefont {D.}~\bibnamefont
  {Khomskii}},\ }\href@noop {} {\emph {\bibinfo {title} {Transition metal
  compounds}}}\ (\bibinfo  {publisher} {Cambridge University Press},\ \bibinfo
  {year} {2014})\BibitemShut {NoStop}%
\bibitem [{\citenamefont {Johannes}\ and\ \citenamefont
  {Mazin}(2008)}]{Mazin_cdw_phonon}%
  \BibitemOpen
  \bibfield  {author} {\bibinfo {author} {\bibfnamefont {M.~D.}\ \bibnamefont
  {Johannes}}\ and\ \bibinfo {author} {\bibfnamefont {I.~I.}\ \bibnamefont
  {Mazin}},\ }\href {\doibase 10.1103/PhysRevB.77.165135} {\bibfield  {journal}
  {\bibinfo  {journal} {Phys. Rev. B}\ }\textbf {\bibinfo {volume} {77}},\
  \bibinfo {pages} {165135} (\bibinfo {year} {2008})}\BibitemShut {NoStop}%
\bibitem [{\citenamefont {Williams}\ \emph {et~al.}(1976)\citenamefont
  {Williams}, \citenamefont {Scruby}, \citenamefont {Clark},\ and\
  \citenamefont {Parry}}]{williams_charge_1976}%
  \BibitemOpen
  \bibfield  {author} {\bibinfo {author} {\bibfnamefont {P.~M.}\ \bibnamefont
  {Williams}}, \bibinfo {author} {\bibfnamefont {C.~B.}\ \bibnamefont
  {Scruby}}, \bibinfo {author} {\bibfnamefont {W.~B.}\ \bibnamefont {Clark}}, \
  and\ \bibinfo {author} {\bibfnamefont {G.~S.}\ \bibnamefont {Parry}},\
  }\href@noop {} {\bibfield  {journal} {\bibinfo  {journal} {Le Journal de
  Physique Colloques}\ }\textbf {\bibinfo {volume} {37}},\ \bibinfo {pages}
  {C4} (\bibinfo {year} {1976})}\BibitemShut {NoStop}%
\bibitem [{\citenamefont {Lu}\ \emph {et~al.}(2017)\citenamefont {Lu},
  \citenamefont {Zhu}, \citenamefont {Xiao}, \citenamefont {Chuu},
  \citenamefont {Han}, \citenamefont {Chiu}, \citenamefont {Cheng},
  \citenamefont {Yang}, \citenamefont {Wei}, \citenamefont {Yang},\ and\
  \citenamefont {et~al.}}]{Lu2017}%
  \BibitemOpen
  \bibfield  {author} {\bibinfo {author} {\bibfnamefont {A.-Y.}\ \bibnamefont
  {Lu}}, \bibinfo {author} {\bibfnamefont {H.}~\bibnamefont {Zhu}}, \bibinfo
  {author} {\bibfnamefont {J.}~\bibnamefont {Xiao}}, \bibinfo {author}
  {\bibfnamefont {C.-P.}\ \bibnamefont {Chuu}}, \bibinfo {author}
  {\bibfnamefont {Y.}~\bibnamefont {Han}}, \bibinfo {author} {\bibfnamefont
  {M.-H.}\ \bibnamefont {Chiu}}, \bibinfo {author} {\bibfnamefont {C.-C.}\
  \bibnamefont {Cheng}}, \bibinfo {author} {\bibfnamefont {C.-W.}\ \bibnamefont
  {Yang}}, \bibinfo {author} {\bibfnamefont {K.-H.}\ \bibnamefont {Wei}},
  \bibinfo {author} {\bibfnamefont {Y.}~\bibnamefont {Yang}}, \ and\ \bibinfo
  {author} {\bibnamefont {et~al.}},\ }\href {\doibase 10.1038/nnano.2017.100}
  {\bibfield  {journal} {\bibinfo  {journal} {Nature Nanotechnology}\ }\textbf
  {\bibinfo {volume} {12}},\ \bibinfo {pages} {744} (\bibinfo {year}
  {2017})}\BibitemShut {NoStop}%
\bibitem [{\citenamefont {Maghirang}\ \emph {et~al.}(2019)\citenamefont
  {Maghirang}, \citenamefont {Huang}, \citenamefont {Villaos}, \citenamefont
  {Hsu}, \citenamefont {Feng}, \citenamefont {Florido}, \citenamefont {Lin},
  \citenamefont {Bansil},\ and\ \citenamefont {Chuang}}]{Maghirang2019}%
  \BibitemOpen
  \bibfield  {author} {\bibinfo {author} {\bibfnamefont {A.~B.}\ \bibnamefont
  {Maghirang}}, \bibinfo {author} {\bibfnamefont {Z.-Q.}\ \bibnamefont
  {Huang}}, \bibinfo {author} {\bibfnamefont {R.~A.~B.}\ \bibnamefont
  {Villaos}}, \bibinfo {author} {\bibfnamefont {C.-H.}\ \bibnamefont {Hsu}},
  \bibinfo {author} {\bibfnamefont {L.-Y.}\ \bibnamefont {Feng}}, \bibinfo
  {author} {\bibfnamefont {E.}~\bibnamefont {Florido}}, \bibinfo {author}
  {\bibfnamefont {H.}~\bibnamefont {Lin}}, \bibinfo {author} {\bibfnamefont
  {A.}~\bibnamefont {Bansil}}, \ and\ \bibinfo {author} {\bibfnamefont {F.-C.}\
  \bibnamefont {Chuang}},\ }\href {\doibase 10.1038/s41699-019-0118-2}
  {\bibfield  {journal} {\bibinfo  {journal} {npj 2D Materials and
  Applications}\ }\textbf {\bibinfo {volume} {3}},\ \bibinfo {pages} {35}
  (\bibinfo {year} {2019})}\BibitemShut {NoStop}%
\bibitem [{\citenamefont {Terashima}\ \emph {et~al.}(2003)\citenamefont
  {Terashima}, \citenamefont {Sato}, \citenamefont {Komatsu}, \citenamefont
  {Takahashi}, \citenamefont {Maeda},\ and\ \citenamefont
  {Hayashi}}]{terashima_charge-density_2003}%
  \BibitemOpen
  \bibfield  {author} {\bibinfo {author} {\bibfnamefont {K.}~\bibnamefont
  {Terashima}}, \bibinfo {author} {\bibfnamefont {T.}~\bibnamefont {Sato}},
  \bibinfo {author} {\bibfnamefont {H.}~\bibnamefont {Komatsu}}, \bibinfo
  {author} {\bibfnamefont {T.}~\bibnamefont {Takahashi}}, \bibinfo {author}
  {\bibfnamefont {N.}~\bibnamefont {Maeda}}, \ and\ \bibinfo {author}
  {\bibfnamefont {K.}~\bibnamefont {Hayashi}},\ }\href {\doibase
  10.1103/PhysRevB.68.155108} {\bibfield  {journal} {\bibinfo  {journal}
  {Physical Review B}\ }\textbf {\bibinfo {volume} {68}} (\bibinfo {year}
  {2003}),\ 10.1103/PhysRevB.68.155108}\BibitemShut {NoStop}%
\bibitem [{\citenamefont {Borisenko}\ \emph {et~al.}(2008)\citenamefont
  {Borisenko}, \citenamefont {Kordyuk}, \citenamefont {Yaresko}, \citenamefont
  {Zabolotnyy}, \citenamefont {Inosov}, \citenamefont {Schuster}, \citenamefont
  {B\"uchner}, \citenamefont {Weber}, \citenamefont {Follath}, \citenamefont
  {Patthey},\ and\ \citenamefont {et~al.}}]{PhysRevLett.100.196402}%
  \BibitemOpen
  \bibfield  {author} {\bibinfo {author} {\bibfnamefont {S.~V.}\ \bibnamefont
  {Borisenko}}, \bibinfo {author} {\bibfnamefont {A.~A.}\ \bibnamefont
  {Kordyuk}}, \bibinfo {author} {\bibfnamefont {A.~N.}\ \bibnamefont
  {Yaresko}}, \bibinfo {author} {\bibfnamefont {V.~B.}\ \bibnamefont
  {Zabolotnyy}}, \bibinfo {author} {\bibfnamefont {D.~S.}\ \bibnamefont
  {Inosov}}, \bibinfo {author} {\bibfnamefont {R.}~\bibnamefont {Schuster}},
  \bibinfo {author} {\bibfnamefont {B.}~\bibnamefont {B\"uchner}}, \bibinfo
  {author} {\bibfnamefont {R.}~\bibnamefont {Weber}}, \bibinfo {author}
  {\bibfnamefont {R.}~\bibnamefont {Follath}}, \bibinfo {author} {\bibfnamefont
  {L.}~\bibnamefont {Patthey}}, \ and\ \bibinfo {author} {\bibnamefont
  {et~al.}},\ }\href {\doibase 10.1103/PhysRevLett.100.196402} {\bibfield
  {journal} {\bibinfo  {journal} {Phys. Rev. Lett.}\ }\textbf {\bibinfo
  {volume} {100}},\ \bibinfo {pages} {196402} (\bibinfo {year}
  {2008})}\BibitemShut {NoStop}%
\bibitem [{\citenamefont {Wang}\ \emph {et~al.}(2011)\citenamefont {Wang},
  \citenamefont {Eyert},\ and\ \citenamefont {Schwingenschlögl}}]{Wang_2011}%
  \BibitemOpen
  \bibfield  {author} {\bibinfo {author} {\bibfnamefont {H.}~\bibnamefont
  {Wang}}, \bibinfo {author} {\bibfnamefont {V.}~\bibnamefont {Eyert}}, \ and\
  \bibinfo {author} {\bibfnamefont {U.}~\bibnamefont {Schwingenschlögl}},\
  }\href {\doibase 10.1088/0953-8984/23/11/116003} {\bibfield  {journal}
  {\bibinfo  {journal} {Journal of Physics: Condensed Matter}\ }\textbf
  {\bibinfo {volume} {23}},\ \bibinfo {pages} {116003} (\bibinfo {year}
  {2011})}\BibitemShut {NoStop}%
\bibitem [{\citenamefont {O.~Fumega}\ \emph {et~al.}(2020)\citenamefont
  {O.~Fumega}, \citenamefont {Blanco-Canosa}, \citenamefont {Babu-Vasili},
  \citenamefont {Gargiani}, \citenamefont {Li}, \citenamefont {Zhou},
  \citenamefont {Rivadulla},\ and\ \citenamefont {Pardo}}]{OFumega2020}%
  \BibitemOpen
  \bibfield  {author} {\bibinfo {author} {\bibfnamefont {A.}~\bibnamefont
  {O.~Fumega}}, \bibinfo {author} {\bibfnamefont {S.}~\bibnamefont
  {Blanco-Canosa}}, \bibinfo {author} {\bibfnamefont {H.}~\bibnamefont
  {Babu-Vasili}}, \bibinfo {author} {\bibfnamefont {P.}~\bibnamefont
  {Gargiani}}, \bibinfo {author} {\bibfnamefont {H.}~\bibnamefont {Li}},
  \bibinfo {author} {\bibfnamefont {J.-S.}\ \bibnamefont {Zhou}}, \bibinfo
  {author} {\bibfnamefont {F.}~\bibnamefont {Rivadulla}}, \ and\ \bibinfo
  {author} {\bibfnamefont {V.}~\bibnamefont {Pardo}},\ }\href {\doibase
  10.1039/D0TC02003F} {\bibfield  {journal} {\bibinfo  {journal} {Journal of
  Materials Chemistry C}\ }\textbf {\bibinfo {volume} {in press}} (\bibinfo
  {year} {2020}),\ 10.1039/D0TC02003F}\BibitemShut {NoStop}%
\bibitem [{\citenamefont {Lin}\ \emph {et~al.}(2018)\citenamefont {Lin},
  \citenamefont {Lohmann}, \citenamefont {Ali}, \citenamefont {Tang},
  \citenamefont {Li}, \citenamefont {Xing}, \citenamefont {Zhong},
  \citenamefont {Jia}, \citenamefont {Han}, \citenamefont {Coh},\ and\
  \citenamefont {et~al.}}]{spin_anys_press}%
  \BibitemOpen
  \bibfield  {author} {\bibinfo {author} {\bibfnamefont {Z.}~\bibnamefont
  {Lin}}, \bibinfo {author} {\bibfnamefont {M.}~\bibnamefont {Lohmann}},
  \bibinfo {author} {\bibfnamefont {Z.~A.}\ \bibnamefont {Ali}}, \bibinfo
  {author} {\bibfnamefont {C.}~\bibnamefont {Tang}}, \bibinfo {author}
  {\bibfnamefont {J.}~\bibnamefont {Li}}, \bibinfo {author} {\bibfnamefont
  {W.}~\bibnamefont {Xing}}, \bibinfo {author} {\bibfnamefont {J.}~\bibnamefont
  {Zhong}}, \bibinfo {author} {\bibfnamefont {S.}~\bibnamefont {Jia}}, \bibinfo
  {author} {\bibfnamefont {W.}~\bibnamefont {Han}}, \bibinfo {author}
  {\bibfnamefont {S.}~\bibnamefont {Coh}}, \ and\ \bibinfo {author}
  {\bibnamefont {et~al.}},\ }\href {\doibase 10.1103/PhysRevMaterials.2.051004}
  {\bibfield  {journal} {\bibinfo  {journal} {Phys. Rev. Materials}\ }\textbf
  {\bibinfo {volume} {2}},\ \bibinfo {pages} {051004} (\bibinfo {year}
  {2018})}\BibitemShut {NoStop}%
\bibitem [{\citenamefont {Nova}\ \emph {et~al.}(2019)\citenamefont {Nova},
  \citenamefont {Disa}, \citenamefont {Fechner},\ and\ \citenamefont
  {Cavalleri}}]{Nova1075}%
  \BibitemOpen
  \bibfield  {author} {\bibinfo {author} {\bibfnamefont {T.~F.}\ \bibnamefont
  {Nova}}, \bibinfo {author} {\bibfnamefont {A.~S.}\ \bibnamefont {Disa}},
  \bibinfo {author} {\bibfnamefont {M.}~\bibnamefont {Fechner}}, \ and\
  \bibinfo {author} {\bibfnamefont {A.}~\bibnamefont {Cavalleri}},\ }\href
  {\doibase 10.1126/science.aaw4911} {\bibfield  {journal} {\bibinfo  {journal}
  {Science}\ }\textbf {\bibinfo {volume} {364}},\ \bibinfo {pages} {1075}
  (\bibinfo {year} {2019})}\BibitemShut {NoStop}%
\end{thebibliography}
\end{document}